\documentclass[pra,showpacs,twocolumn,superscriptaddress]{revtex4}

\newcommand{\ket}[1]{|{#1}\rangle}

\newcommand{\Melement}[3]{\left\langle #1 \right| #3 \left| #2 \right\rangle } 

\usepackage{amsfonts}
\usepackage{amsmath}
\usepackage{graphicx}
\usepackage{amssymb}
\usepackage{color}
\usepackage{subfigure}

\usepackage{mathtools}

\begin{document}
\title{Vortex dynamics in superfluids governed by an interacting gauge theory}

\author{Salvatore Butera}
\author{Manuel Valiente}
\author{Patrik \"Ohberg}
\affiliation{SUPA, Institute of Photonics and Quantum Sciences, Heriot-Watt University, Edinburgh EH14 4AS, United Kingdom}
\begin{abstract}
We study the dynamics of a vortex in a quasi two-dimensional Bose gas consisting of light matter coupled atoms forming two-component pseudo spins. The gas is subject to a density dependent gauge potential, hence governed by an interacting gauge theory, which stems from a collisionally induced detuning between the incident laser frequency and the atomic energy levels. This provides a back-action between the synthetic gauge potential and the matter field.  A Lagrangian approach is used to derive an expression for the force acting on a vortex in such a gas.  We discuss the similarities between this force and the one predicted by Iordanskii, Lifshitz and Pitaevskii when scattering between a superfluid vortex and the thermal component is taken into account.
\end{abstract}
\maketitle

\section{Introduction}
The interest in vortex states, and more generally in the rotational properties of fluids, dates back to the early days of hydrodynamics and is historically related to the phenomenon of turbulence in classical fluids. To understand the onset of chaotic dynamics and turbulence has turned out to be a formidable task, and a deep and complete understanding is still far from achieved.
The realization of Bose-Einstein condensation of $\prescript{4}{}{\text{He}}$ \cite{kapitza_1938,allen_1938}, and the consequent discovery of superfluidity opened up a new perspective to this aim. As in their classical counterpart, turbulence also shows up in these quantum fluids, with vortices playing a central role in the transition to chaotic motion \cite{Gorter_1949,Feynman_1955,Vinen_1957a,Vinen_1957b,Vinen_1957c}.

The advantage of investigating turbulence phenomena in superfluids is due to the constraints that quantum mechanics imposes on the values of the physical quantities that characterize the system, which simplifies to some extent the scenario with respect to its classical counterpart. For example, in order for the condensate wave function to be single valued, the circulation of the velocity field around any closed path, has to be quantized in multiples of $\hbar/m$, with $m$ the mass of the atomic species composing the condensate itself. This property leads to the concept of quantized vortices, around which the circulation (and the angular momentum as a consequence) is quantized \cite{Onsager_1949,Feynman_1955}. Apart from the discreteness of the values of the angular momentum, a vortex in a superfluid has the remarkable property of being a particle-like stable object that does not easily decay, in contrast to viscous diffusion of vorticity, as in the case of classical fluids. Because of these considerations, superfluids have become the preferred environment for investigating turbulence phenomena. The study of the dynamics of quantized vortices represented the first step to this aim. The experimental realization of Bose-Einstein condensation (BEC) of alkali atoms in 1995 \cite{Anderson1995,Davis1995a}, gave a significant boost in this direction. Because of the unprecedented control and access to physical parameters of the atomic cloud, these systems have provided an excellent experimental environment for studying the dynamics of quantized vortices and their properties in general \cite{Matthews1999a,Madison2000,lin_2009b}.

Particular attention has been drawn to the problem of Magnus like transverse forces in quantum fluids. These forces, first predicted in classical hydrodynamics, are orthogonal to the relative motion between an object, carrying a flow of circulation, and the fluid in which it is immersed. The forces at play in this situation, and their derivation, has not been without controversy. At the heart of this debate,  is the dual nature of a quantum fluid at finite temperature, where it consists of a superfluid (condensed) and a normal component of thermally excited quasi-particles. According to this two-fluid model, different transverse forces acting on a vortex should in principle be expected. Whereas there is a wide consensus on the existence of a superfluid Magnus force, which can be considered the analogue of the effect predicted by the Kutta-Joukowski theorem for an inviscid classical fluid, the existence of a thermal Magnus force is still the object of some debate. Such forces was theoretically predicted by Lifshitz and Pitaevskii  \cite{Lifshitz_1958} and  Iordanskii \cite{Iordansky_1964,Iordanskii_1966}, who showed that this type of force is a consequence of the interaction between a vortex and the roton and phonon quasi-particles respectively.

Recently  Ao and Thouless \cite{Thouless_1996,Thouless_1993} contested the existence of any thermal Magnus force. Deircan et al \cite{Demircan_1995} arrived at the same conclusion analysing the phonon scattering by a vortex using a hydrodynamical approach. These results have been confuted by Sonin \cite{Sonin_1975,Sonin_1976,Sonin_1997}, who argued  it is incorrect ignoring particular properties of the Born cross-section at small angles, which if included, results in a thermal transverse force. However, despite all the efforts and theoretical work done on the subject, a clear and definitive conclusion about the existence of these thermal Magnus forces still remain a topic of debate.

In this paper we will study the motion of a vortex in a superfluid which is subject to a density dependent gauge potential. We will show that the resulting force on the vortex is similar to the Iordanskii force. In the quest to find a physical system which would emulate a dynamical gauge theory, it has been proposed, as a first step towards such a situation, to use collisionally induced detunings in combination with synthetic magnetism arising from light-matter coupling \cite{dalibard_2011,goldman_2013}. The resulting gauge field is not dynamical in a field theoretic sense, but it does become density dependent and therefore provides a back-action between the synthetic gauge potential and the superfluid. This results in a current nonlinearity in the equation of motion for the superfluid with dramatic consequences for the transport properties of the system \cite{edmonds_2013a,greschner_2013,zheng_2014}. A more complete understanding of the vortex dynamics in such a system will provide important insight into phenomena such as drag forces and superfluidity of the chiral gas.

We start by briefly introducing the concept of a synthetic nonlinear gauge potential. Following a variational approach, we study the dynamics of the vortex core, explicitly calculating the forces acting on it. We finally validate our arguments by comparing the analytical results with a numerical solution of the generalised Gross-Pitaevskii equation.

\section{Atoms in artificial gauge fields}

\begin{figure}[h]
\includegraphics[width=\linewidth]{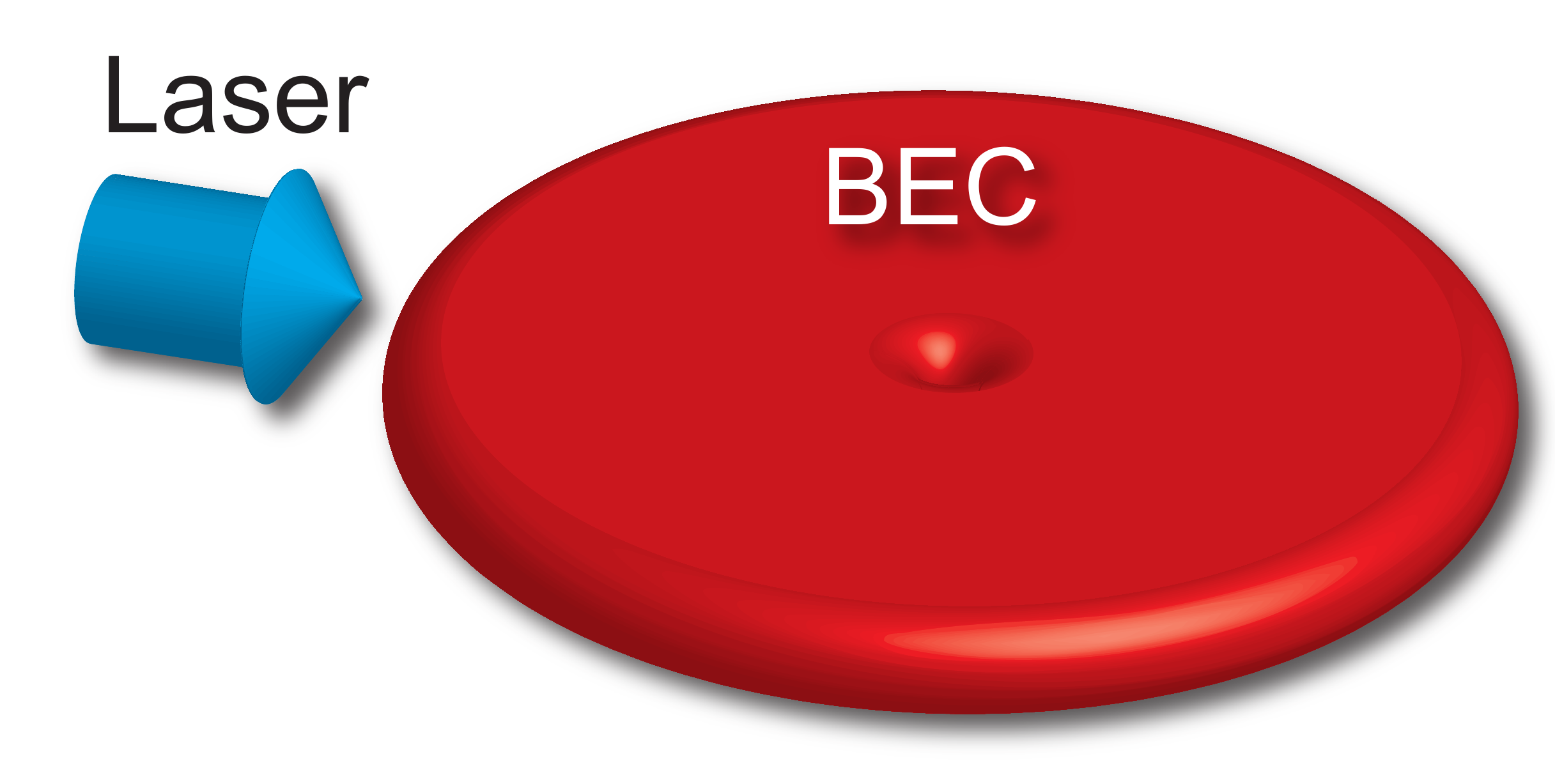}
\caption{Schematic view of the experimental setup with a vortex located in a Bose-Einstein condensate and a laser beam incident in the plane of the quasi two-dimensional condensate.}
\label{atom}
\end{figure}

We consider a Bose Einstein condensate consisting of two-level atoms, confined in a highly anisotropic trap, such that its dynamics in the transverse direction is frozen, and the system can be considered as a quasi-two-dimensional cloud of atoms, nearly free to move in a plane. 
It has recently been shown in \cite{edmonds_2013a} that a collisionally induced detuning between the incident laser and the two atomic levels can give rise to a density dependent synthetic gauge potential. We refer the reader to  Appendix A for a detailed derivation of the equation of motion. The resulting mean field equation which describes the dynamics of the condensate is given by
\begin{equation}
	i\hbar\frac{\partial\psi}{\partial t}=\left[\frac{\left(\mathbf{p}-\mathbf{A}\right)^2}{2m}-\mathbf{a}_1\cdot\mathbf{j}+W+g\rho\right]\psi
\label{GP1}
\end{equation}
in which an unconventional nonlinearity appears, proportional to the current
\begin{equation}
	\mathbf{j}=\frac{\hbar}{2mi}\left[\psi^*\left(\nabla-\frac{i}{\hbar}\mathbf{A}\right)\psi-\psi\left(\nabla+\frac{i}{\hbar}\mathbf{A}\right)\psi^*\right].
\label{Current}
\end{equation}
The validity of Eq.  (\ref{GP1}) relies on the adiabatic approximation where the atoms are assumed to be prepared in one of the dressed states of the light-matter coupled system, and on the assumption that the Rabi frequency is the dominating energy scale. The resulting gauge potential and scalar potential are given by
\begin{eqnarray}
	\mathbf{A}& =&\mathbf{A}^{(0)}\pm \mathbf{a}_1 \rho(\mathbf{r})\\
	W &=&\frac{\left| \mathbf{A}^{(0)}\right|^2}{2m}.
\label{Potentials_def}
\end{eqnarray}
Here, and through the rest of the paper, $\pm$ refers to the two different dressed states which we can choose to use (see Appendix A). The  $\mathbf{A}^{(0)}=-\frac{\hbar}{2}\nabla\phi$ is the single particle component of the vector potential with $\phi$ being the phase of the laser. The vector field $\mathbf{a}_1=\nabla\phi\,(g_{11}-g_{22})/8\Omega$ is the first order nonlinear density dependent contribution where $g_{11}$ and $g_{22}$ are the corresponding meanfield coupling constants for collisions between atoms in state $|1\rangle$ and $|2\rangle$ respectively, and $\Omega$ is the Rabi frequency. See Fig.  \ref{atom} for a description of the envisaged setup. 

\section{Vortex Lagrangian}

We consider a cloud which is strongly confined by a potential in the $z$ direction, such that the dynamics is frozen in this direction and the atoms are nearly free to move in the x-y plane. We assume an incident laser beam which is propagating in the plane of the condensate and with a uniform intensity and phase $\phi(\mathbf{r})=\mathbf{k}\cdot\mathbf{r}$. The Rabi frequency is consequently uniform throughout the condensate. This choice of light beam results in zero magnetic field, but the nonlinear part of the gauge potential will, as we show next, influence the dynamics. The zeroth order gauge potential $\mathbf{A}^{(0)}$ can be gauged away by applying the transformation $\Psi\rightarrow \exp\left(\mp i\frac{\mathbf{k}\cdot\mathbf{r}}{2}\right) \Psi$ which results in 
\begin{align}
	\mathbf{A}& =\pm\mathbf{a}_1 \rho(\mathbf{r})\label{Vec_pot}\\
	W & =\frac{\hbar^2 k^2}{8m}.\label{Scal_pot}
\end{align}
with $\mathbf{a}_1=\mathbf{k}(g_{11}-g_{22})/\left(8\Omega\right)$. In order to study the dynamics of the vortex in the cloud, it is convenient to consider the cloud having an effective thickness $Z$ in the $z$-direction. The original condensate wave function can then be rescaled as $\psi(\mathbf{r})/\sqrt{Z}$, where $\psi(\mathbf{r})$ is now two-dimensional, and normalized in such a way that $\int{d^2\mathbf{r}|\psi|^2}=N$, with $N$ the number of atoms in the condensate. 

We write the wave function in terms of the particle density $\rho$ and the phase $S$ as $\psi=\sqrt{\rho}e^{iS}$, so that the Lagrangian density takes the form
\begin{eqnarray}
\mathcal{L}&=&-\hbar\rho\frac{\partial S}{\partial t}-\frac{\hbar^2}{2m}\left(\nabla\sqrt{\rho}\right)^2-\frac{1}{2}m\rho u^2\nonumber\\
&&-\frac{i\hbar}{2m}\rho\left(\nabla\cdot\mathbf{A}\right)-\frac{i\hbar}{m}\sqrt{\rho}\,\mathbf{A}\cdot\left(\nabla\sqrt{\rho}\right)\nonumber\\
&&+\frac{h}{m}\rho\left(\mathbf{A}\cdot\nabla S\right)-\rho\left[W+\frac{\hbar\Omega}{2}+\frac{g}{2}\rho\right]
\label{Lagr}
\end{eqnarray}
where the physical velocity $\mathbf{u}$ in the condensate is related to the phase of the wave function as
\begin{equation}
	m \mathbf{u}=\hbar\nabla S-\mathbf{A}.
\label{u_S_A}
\end{equation}
Given Eq.  \eqref{Lagr}, we seek for an effective Lagrangian which describes the dynamics of a vortex state. We look in particular for the forces which result from the vortex interacting with the synthetic gauge field. 
In order to properly take into account the vortex velocity field we need to choose the phase $S$, in such a way that $\hbar\nabla S=m \mathbf{u}_0(\mathbf{r})$, with
\begin{equation}
	\mathbf{u}_0(\mathbf{r)}=\frac{\boldsymbol{\kappa}\times\mathbf{r}}{2\pi r^2}
\label{u_vortex}
\end{equation}
the velocity field characteristic of the vortex state, and $\kappa=|\boldsymbol{\kappa}|=h/m$ the quantum of circulation. From an experimental point of view this is equivalent to preparing a vortex in the atomic cloud in absence of the gauge potentials, and then look at its dynamics once the external laser field is switched on. We next consider the vortex moving relative to the bulk condensate, where we indicate by $\mathbf{r}_0$ the position of its core and by $\mathbf{v}=d\mathbf{r}_0/dt$ its velocity. We assume this velocity is much smaller than the speed of sound in the condensate, so that the density and phase profiles charactering the vortex, adiabatically follow the core during its motion without undergoing any distortion. We make the ansatz $\rho=\rho_0(\mathbf{r}-\mathbf{r}_0)$ for the density of the condensate, with $\rho_0(\mathbf{r})$ the density profile of a vortex state, which is assumed to  carry a single quantum of circulation. We write the phase of the condensate as $S=S_0+S_v$ with $S_0$ the phase of a steady vortex, so that $\nabla S_0=\mathbf{u}_0$, and $S_v$ the shift due to the core's motion. Exploiting the continuity equation
\begin{equation}
	\frac{d\rho}{dt}+\nabla\cdot(\rho\mathbf{u}) =0
\label{continuity}
\end{equation}
we get the  equation for $S_v$. To do so we substitute Eq. \eqref{u_S_A} into Eq. \eqref{continuity}, obtaining
\begin{equation}
	\nabla\rho\cdot\left(\hbar\nabla S_v-\mathbf{A}-m\mathbf{V}\right)+\rho\left(\hbar\Delta S_v-\nabla\cdot\mathbf{A} \right)
\label{Sv_1}
\end{equation}
because $\nabla\rho\cdot\nabla S_0=0$ and $\Delta S_0\equiv\nabla\cdot\nabla S_0=0$. The vortex is assumed to be moving in the condensate with constant velocity. We therefore expect that $\nabla S_v$ gives rise to an uniform field, so that $\Delta S_v=0$, and Eq. \eqref{Sv_1} reduces to
\begin{equation}
	\nabla\rho\cdot\left(\hbar\nabla S_v-m\mathbf{v}\pm 2\rho\mathbf{a}_1\right)=0
\label{Sv}
\end{equation}
having used the relation $\nabla\cdot\left(\rho\mathbf{A}\right)=\pm\nabla\cdot\left(\rho^2\mathbf{a}_1\right)=\pm 2\rho\nabla\rho\cdot\mathbf{a}_1$. It is useful now to distinguish between the in-core (in which $\nabla\rho\neq 0$, $\rho\approx 0$) and out-core (in which $\nabla\rho\approx 0$, $\rho\neq 0$) regions of the vortex. With this distinction in mind, Eq. \eqref{Sv} can be solved, giving
\begin{align}
	\hbar\nabla S_v&=m\mathbf{v} &\text{in-core} \label{S_in_core}\\
	\hbar\nabla S_v&=0 &\text{out-core} \label{S_out_core}
\end{align}
The result in Eq. \eqref{S_in_core}  follows straightforwardly from Eq. \eqref{Sv}. 
In Eq. \eqref{S_out_core} we have chosen the boundary conditions such that the mass current is zero at infinity. In order to take advantage of these results, we need to identify in Eq.  \eqref{Lagr} the terms referring to the different regions of the vortex. To do so, we split terms of the type $\rho f\left(\nabla S\right)$ (with $f(\cdot)$ a generic function) into $\left(\rho-\rho_B\right)f\left(\nabla S\right)+\rho_B f\left(\nabla S\right)$, with $\rho_B$ the bulk density of the condensate. The first term is different from zero within a distance from the core of the order of the healing length of the condensate, defined as $\xi=\hbar/\sqrt{2m\rho_B g}$, and so refers to the in-core region, while the second one is relative to the out-core region. Substituting the expression for $\nabla S_v$ in the different terms, and noticing that $\partial_t S=-\nabla S \cdot\mathbf{v}$, we obtain
\begin{eqnarray}
\mathcal{L}&=& \frac{1}{2}m(\rho-\rho_B)v^2+\left[m\rho_B \mathbf{u}_0+(\rho-\rho_B)\mathbf{A}\right]\cdot\mathbf{v}\nonumber\\
&&+\rho\mathbf{A}\cdot\mathbf{u}_0-\frac{1}{2}m\rho u_0^2-\frac{\hbar^2}{2m}(\nabla\sqrt{\rho})^2\nonumber\\
&&-\frac{i\hbar}{2m}\rho\left(\nabla\cdot\mathbf{A}\right)-\frac{i\hbar}{m}\mathbf{A}\cdot\sqrt{\rho}\left(\nabla\sqrt{\rho}\right).
\label{Lagr_fin}
\end{eqnarray}
Integrating the expression in Eq.  \eqref{Lagr_fin} we get the effective Lagrangian describing the motion of the vortex core, given by
\begin{equation}
	L_v=\int{d^2\mathbf{r} \;\mathcal{L}}=\frac{M_v}{2}v^2+\mathbf{A}_v\cdot\mathbf{v}-U_v
\label{Lagr_core}
\end{equation}
where we defined the effective vortex mass $M_v$ and the effective vector and scalar potentials $A_v$ and $U_v$ as
\begin{align}
	M_v&=m\int{d^2\mathbf{r}\,(\rho_0-\rho_B)} \label{Eff_mass}\\
	A_v&=\int{d^2\mathbf{r}\,m\rho_B\mathbf{u}_0}  \label{Eff_vec}\\
	U_v&=U_0-\int{d^2\mathbf{r}\,\rho_0 \mathbf{A}\cdot\mathbf{u}_0}.  \label{Eff_scal}
\end{align}
The $U_0$ accounts for the remaining terms that do not give any contribution to the vortex dynamics, since their values do not depend on the position of the core $\mathbf{r}_0$. The vortex mass $M_v$ takes a negative  value, and accounts for the missing mass in the condensate due to the presence of the vortex. It diverges logarithmically with the size of the system , and takes the form $M_v=m_{\text{core}} \zeta(L/\xi)$, where  $m_{\text{core}}=-\pi m \rho_B \xi^2/2$ and $\zeta(L/\xi)=4\times\int_0^{L/\xi}{x\left(\rho/\rho_B-1\right)dx}$ is the integral in the dimensionless radial length $x=r/\xi$. For typical atomic clouds $L/\xi$ and $\zeta(L/\xi)$ can take values much larger than one, and increases with the size of the system. For large clouds then, the mass of the vortex can attain a value significantly larger than the core mass. 

\section{Vortex motion}
The Lagrangian in Eq.  \eqref{Lagr_fin} describes the core as a point particle of (negative) mass $M_v$ and positive unit charge, which feels the action of an effective vector potential $A_v$, and a scalar potential $U_v$. We therefore expect there to be two forces at play: a Lorentz-type force $F_M=\mathbf{v}\times\mathbf{B}_v$, with $\mathbf{B}_v=\nabla_0\times\mathbf{A}_v$ the effective magnetic field felt by the core, having defined $\nabla_0\equiv d/d\mathbf{r}_0$, and an electric-type one due to the effect of the scalar potential, and given by $F_I=-\nabla_0 U_v$. In order to determine them explicitly, we start by calculating the effective magnetic field,
\begin{eqnarray}
\mathbf{B}_v&=&\nabla_0\times\mathbf{A}_v\nonumber\\
&=&-m\rho_B\int{d^2\mathbf{r}\,\nabla\times\mathbf{u}_0(\mathbf{r}-\mathbf{r}_0)}\nonumber\\
&=&-m\rho_B\int{d^2\mathbf{r}\,\nabla\times\frac{\boldsymbol{\kappa}\times (\mathbf{r}-\mathbf{r}_0)}{2\pi|\mathbf{r}-\mathbf{r}_0|^2}}\nonumber\\
&=&-m\rho_B\hat{\mathbf{e}}_z\oint{\frac{\boldsymbol{\kappa}\times (\mathbf{r}-\mathbf{r}_0)}{2\pi|\mathbf{r}-\mathbf{r}_0|^2}\cdot d\mathbf{s}}\nonumber\\
&=&-m\rho_B\boldsymbol{\kappa}
\label{Bv}
\end{eqnarray}
from which we obtain
\begin{eqnarray}
F_M&=&\mathbf{v}\times\mathbf{B}_v\nonumber\\
&=&m\rho_B \boldsymbol{\kappa}\times \mathbf{v}.
\label{F_magnus}
\end{eqnarray}
This force is orthogonal to the velocity of the core and physically represent a Magnus effect, as it originates from the relative motion between an object carrying a net vorticity and the condensate bulk. The electric-type force takes instead the form
\begin{eqnarray}
F_I&=&-\nabla_0 U_v\nonumber\\
&=&\nabla_0\left(\int{d^2\mathbf{r}\,\rho_0 \mathbf{A}\cdot\mathbf{u}_0}\right)\nonumber\\
&=&\pm\mathbf{a}_1\times\nabla_0\times\int{d\mathbf{r}^2\rho^2\mathbf{u}_0}\nonumber\\
&=&\mp\mathbf{a}_1\times \hat{\mathbf{e}}_z\oint{\rho^2 (\mathbf{r}-\mathbf{r}_0) \frac{\boldsymbol{\kappa}\times (\mathbf{r}-\mathbf{r}_0)}{2\pi|\mathbf{r}-\mathbf{r}_0|^2}\cdot d\mathbf{s} }\nonumber\\
&=&\pm\rho_B^2\boldsymbol{\kappa}\times\mathbf{a}_1\nonumber\\
&=&\pm\left[\frac{\rho_B \left(g_{11}-g_{22}\right)}{8\hbar\Omega}\right]\rho_B\,\boldsymbol{\kappa}\times\mathbf{p}
\label{F_iordanskii}
\end{eqnarray}
where $\mathbf{p}=\hbar \mathbf{k}$ is the momentum carried by the laser beam.

The expression in Eq.  \eqref{F_iordanskii} has the same form as the Iordanskii transverse force acting on a vortex in a superfluid due to the interaction between the velocity field and a phonon excitation with momentum $\mathbf{p}$ (see Appendix B) and effective particle density $n(\mathbf{p})=\epsilon\rho_B^{\text{3D}}$, with $\rho_B^{\text{3D}}=\rho_B/Z$ the number of particle per unit volume, $\epsilon=\rho_B\left(g_{11}^{\text{3D}}-g_{22}^{\text{3D}}\right)/8\hbar\Omega<1$ the perturbative parameter, playing the role of the particle distribution at momentum $\mathbf{p}$, where $g_{ij}^{\text{3D}}=g_{ij}Z$ is the three-dimensional meanfield coupling constant. There is a significant flexibility in order to emulate this type of transverse forces, because the scattering length difference $a_{11}-a_{22}$ and to some extent the density of the cloud, can be relatively easily changed in an experiment. The magnitude of the wave vector of the laser beam, in analogue to the wave vector of the phonon excitation, is limited by the energy splitting between the two internal states of the atoms constituting the condensate.

\begin{figure}[htbp]
\centering
\includegraphics[width=1\linewidth]{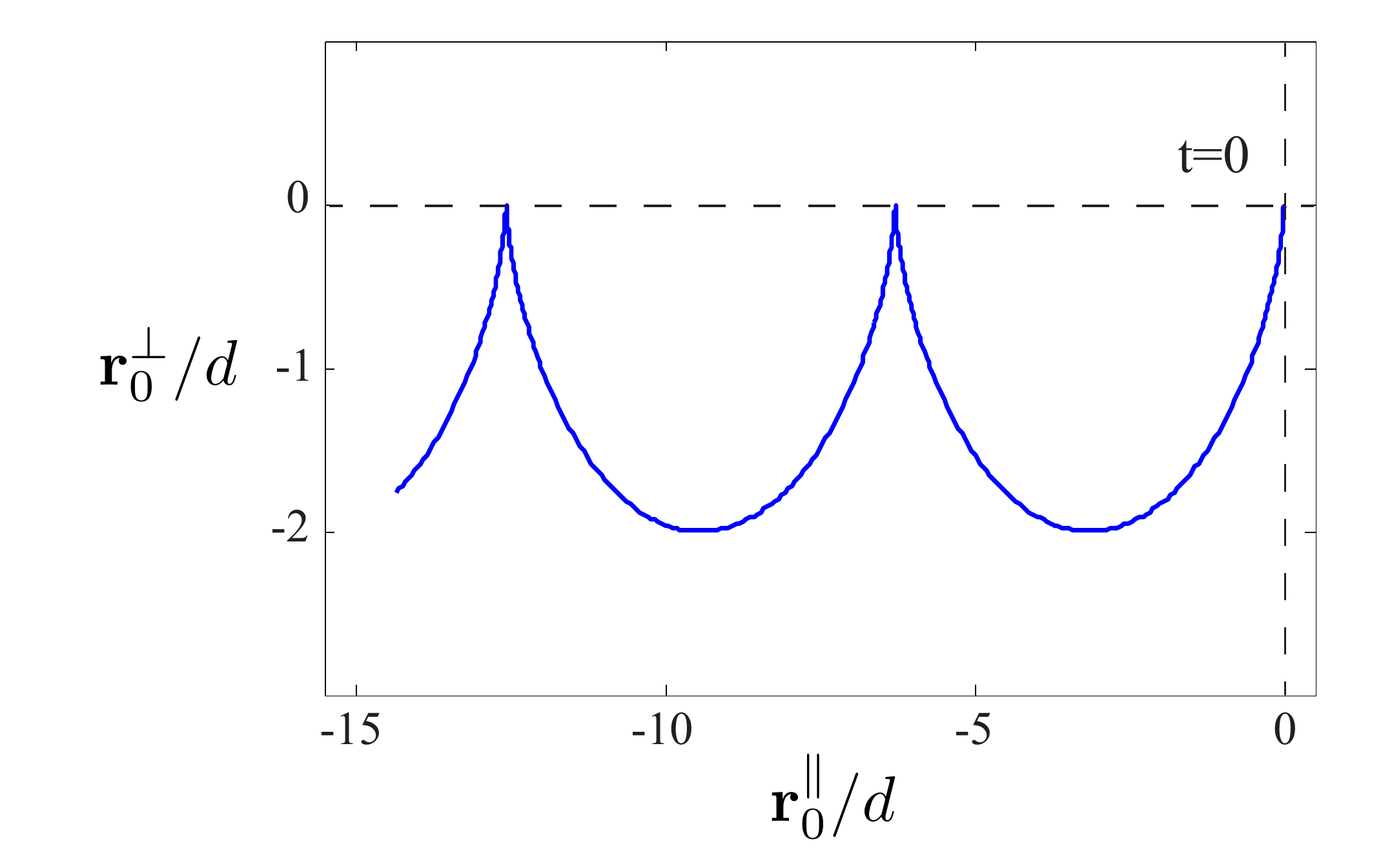}
\caption{Motion of the vortex core given by Eqs.  \eqref{r0_paral} and \eqref{r0_perp} for the + component of the condensate.}
\label{fig:motion}
\end{figure}

With the forces given in Eqs.\eqref{F_magnus} and \eqref{F_iordanskii}, the equation of motion for the vortex core takes the form
\begin{equation}
	M_v\frac{d^2\mathbf{r}_0}{dt^2}=m\rho_B \boldsymbol{\kappa}\times\left(\frac{d\mathbf{r}_0}{dt}\pm\frac{\rho_B}{m}\mathbf{a}_1\right).
\label{Eq_motion}
\end{equation}
For an initially stationary vortex at $\mathbf{r}=0$, the coordinates of the vortex core $\mathbf{r}_0^\parallel$ and $\mathbf{r}_0^\perp$, parallel and orthogonal to the wave vector $\mathbf{k}$ of the laser beam respectively, then becomes
\begin{align}
	\mathbf{r}_0^\parallel&=d\left[\sin\left(\omega t\right)-\omega t\right] \label{r0_paral}\\
	\mathbf{r}_0^\perp&=d\left[\cos\left(\omega t\right)-1\right] \label{r0_perp}
\end{align}
where $d=\pm|M_v||\mathbf{a}_1|/2\pi\hbar m$ and $\omega=2\pi\hbar\rho_B/|M_v|$. Eqs. \eqref{r0_paral} and \eqref{r0_perp} describe a periodic motion for the vortex core, which undergoes a series of curved trajectories of maximum height $d$ and separated by $2\pi d$, as shown in Eqs.   \ref{fig:motion}.

In terms of the healing length, the characteristic length $d$ of the motion, takes the value
\begin{equation}
	\left|\frac{d}{\xi}\right|=\xi\;\frac{\zeta(L)}{4}\epsilon |\mathbf{k}|.
\label{d_xi}
\end{equation}
For typical values of these parameters in atomic clouds with $\xi=0.1 \mu m$, $L/\xi=10-100$, $\lambda=2\pi/|\mathbf{k}|=600\,nm$, and considering a value for the perturbative parameter $\epsilon\sim 0.01$, the ratio between $d$ and the healing length $\xi$ can take values which are significantly larger than one. The characteristic motion of the vortex should therefore be detectable in experiments. 

\begin{figure}[h]
\includegraphics[width=0.8\linewidth]{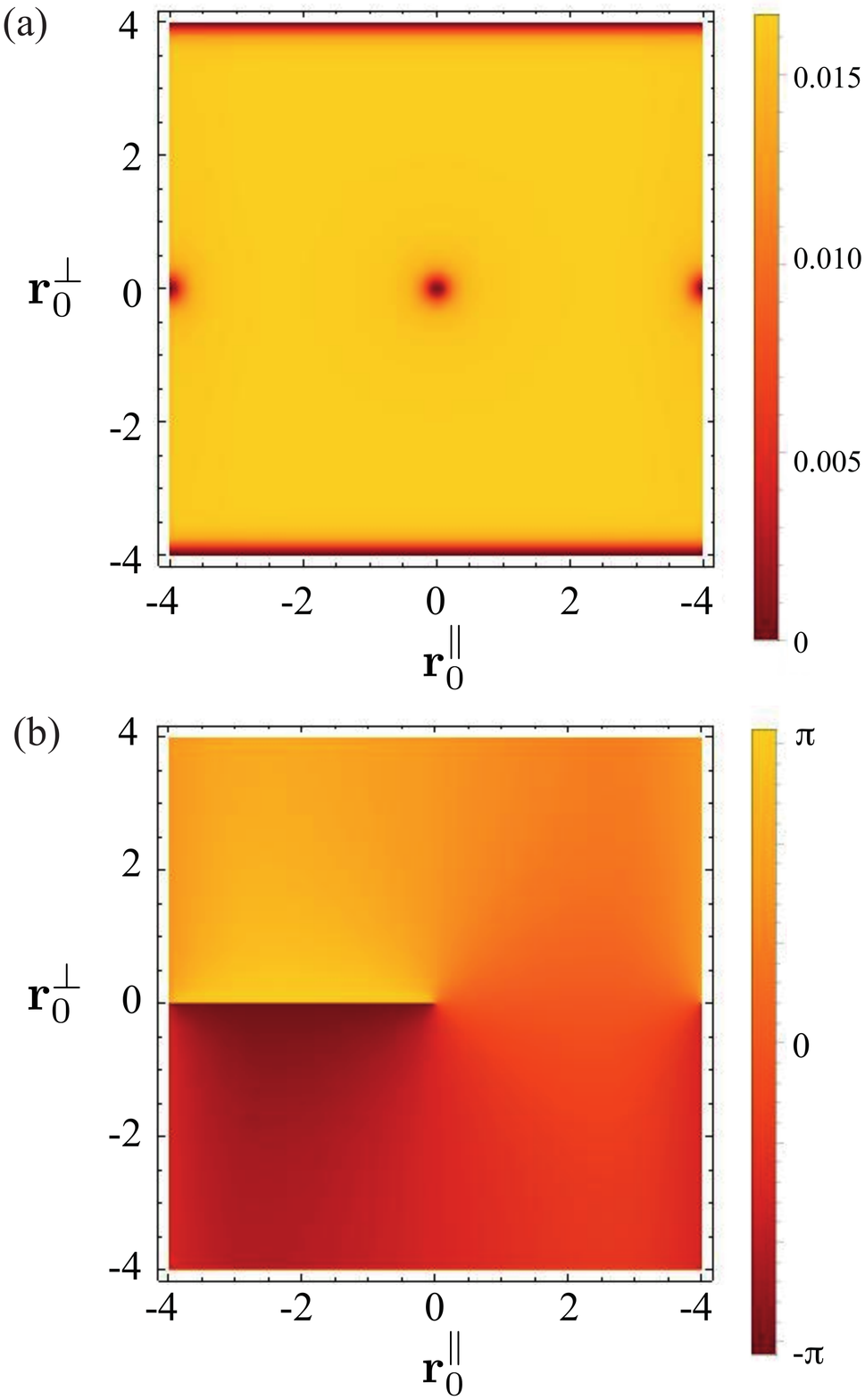}
\caption{Density (a) and phase (b) profiles of the condensate wave function at $t=0$, used as initial condition for the numerical simulation of  the Gross-Pitaevskii equation \eqref{GP}, given by the potentials in Eqs. \eqref{Vec_pot} and \eqref{Scal_pot}. Dimensionless units are used.}
\label{fig:InitCond}
\end{figure}

In order to validate the analytical results, we solved numerically the Gross-Pitaevskii equation \eqref{GP}, with $\mathbf{A}$ and $W$ given in Eqs.  \eqref{Vec_pot} and \eqref{Scal_pot}. We considered a cloud in a square geometry, with periodic boundary conditions in x-direction and confined by a hard-wall potential in the y-direction, giving a homogeneous density which approximates the infinite homogeneous cloud assumed in the analytical description developed above. We determined the initial state of the system by solving Eq.  \eqref{GP} in the imaginary time without the current non-linearity, which leads to the situation represented in Fig. \ref{fig:InitCond}, where two vortices with opposite flow circulation appear in order to match the periodic boundary conditions. Starting from this configuration we compared the numerical simulation with the motion predicted by Eqs. \eqref{r0_paral} and \eqref{r0_perp}. 

\begin{figure}[h]
\centering
\includegraphics[width=\linewidth]{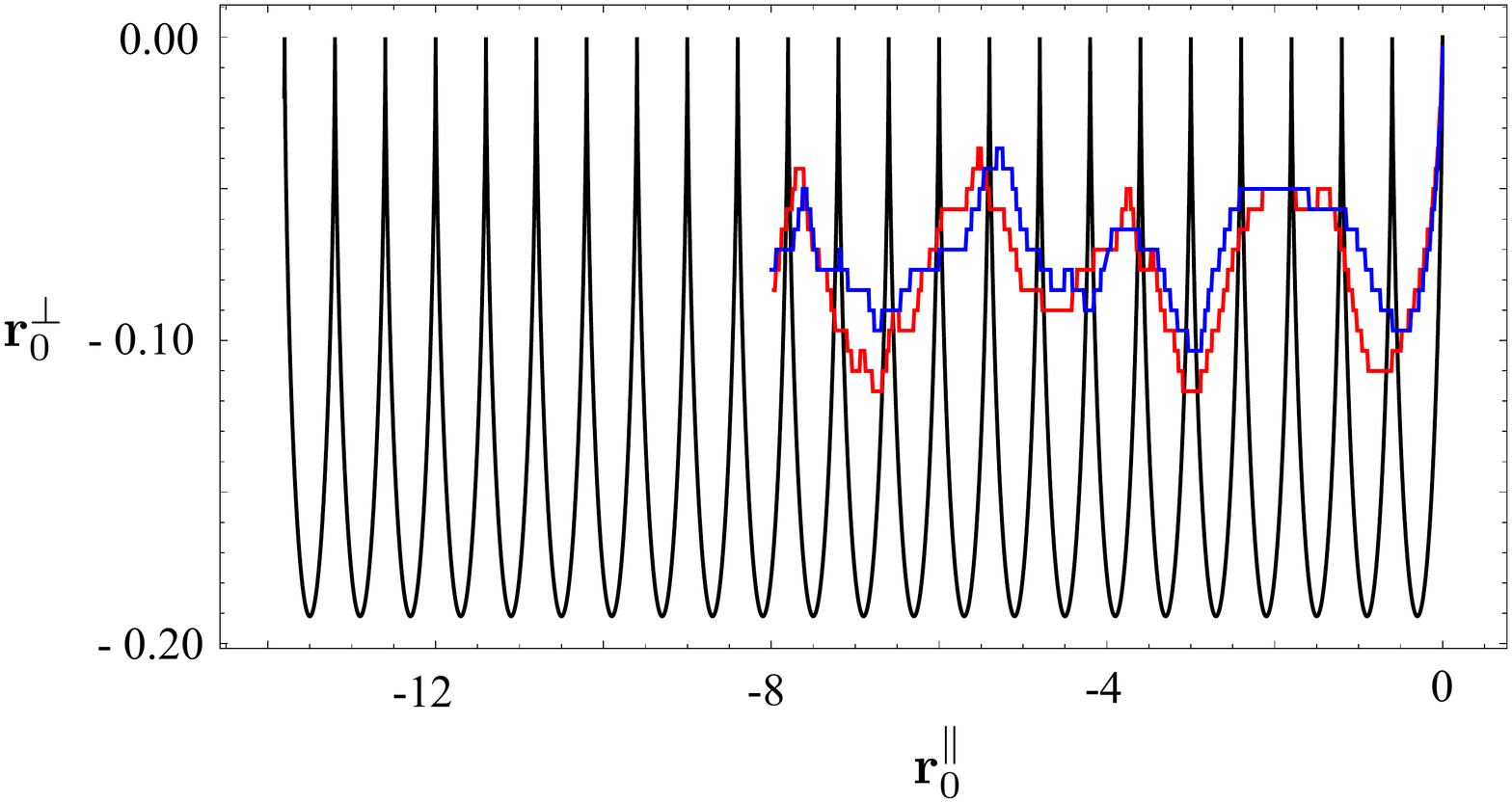}
\caption{Comparison between the analytical (solid black line) and the numerical solution (red and blue curves) for the vortex core motion after $t=2.5$ in units of  $2mL_s^2/\hbar$. The blue curve corresponds to the second vortex in Fig. 3, but plotted as a function of $-{\bf r}_0^\perp$ in order to compare the paths. We see from the numerical curves that the motion of the vortex cores are in opposite direction for  ${\bf r}_0^\perp$ as suggested by Eqs. (\ref{r0_paral}) and (\ref{r0_perp}).}
\label{fig:Match}
\end{figure}

\begin{figure}[h]
\centering
\includegraphics[width=0.8\linewidth]{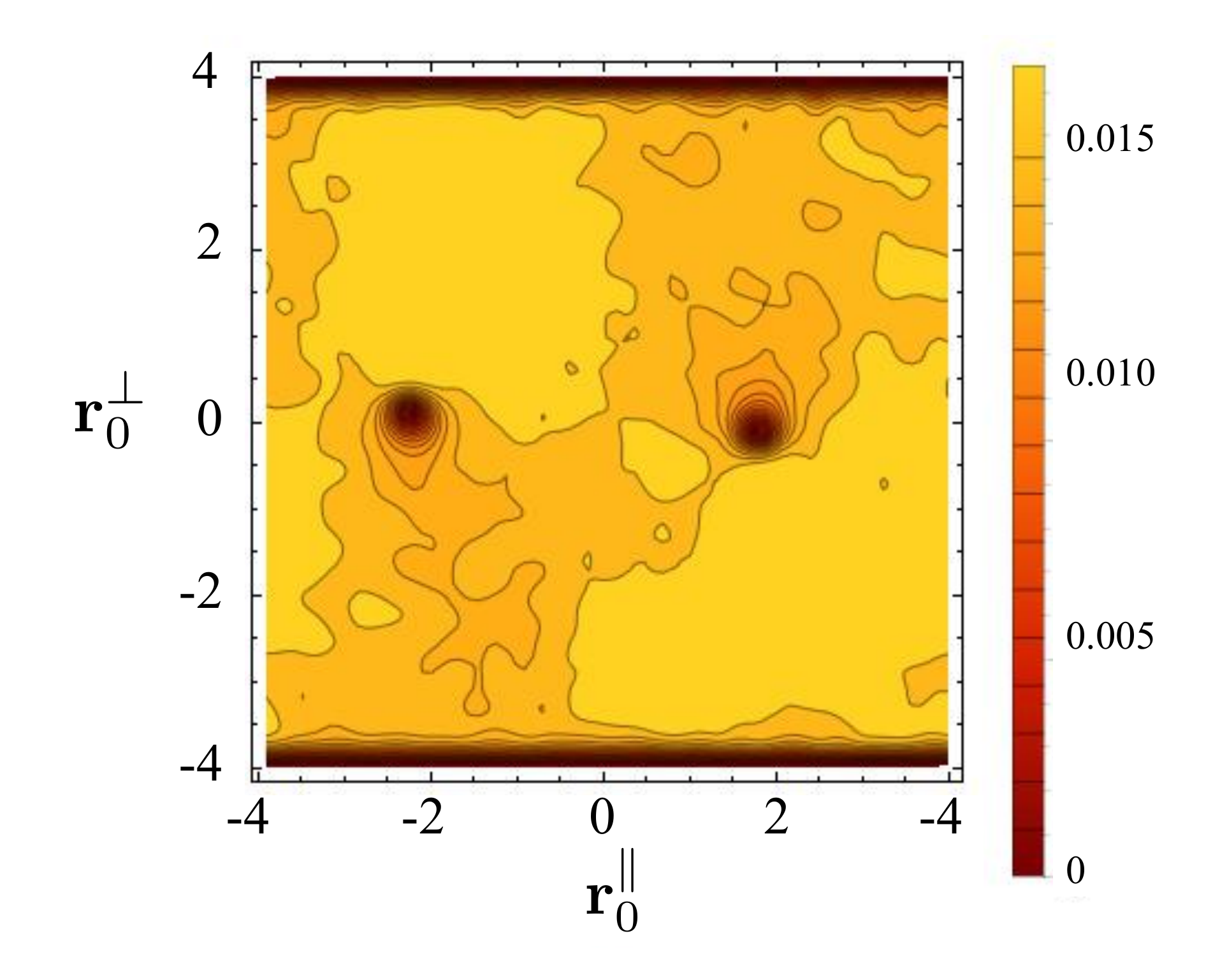}
\caption{Density distribution of the cloud after $t=2.5$ in units of  $2mL_s^2/\hbar$. Deformations in the density are evident compared with  the ideally symmetric ansatz used in the analytical model. As an effect of the current nonlinearity, the meanfield coupling constant takes a different effective value on opposite sides of the vortex core, leading to an asymmetry in the density.}
\label{fig:fluctuations}
\end{figure}

In Fig.  \ref{fig:Match} and Fig.  \ref{fig:fluctuations} we show the dynamics of the vortex core expressed in dimensionless units with $a_1/L_s^2\hbar=0.03$ and $2gm/\hbar^2L_s=1.0$, where $L_s$ is a characteristic length scale, energy is in units of $\hbar^2/2mL_s^2$ and time in units of $2mL_s^2/\hbar$. These parameters can be related to physical values by for instance choosing the atomic mass of Ytterbium, the length $L_s=1\,\mu m$, the combinations of the scattering lengths $a_{11}-a_{22}=65\,nm$ and $(a_{11}+a_{22}+2a_{12})/4=8\,nm$, the Rabi frequency $\Omega=60\,kHz$, the wave length for the incident laser beam to be $\lambda=628\,nm$, and the density of the cloud $3\times10^{14}\,cm^{-3}$ where an effective thickness of the cloud was assumed to be $0.2\,\mu m$. Fig. \ref{fig:Match} shows the numerical simulation for the motion of the vortex core, compared with the analytical solution.  The parameters involved in Eqs. \eqref{r0_paral} and \eqref{r0_perp}, i.e. the bulk density $\rho_B$ of the cloud and the effective vortex mass $M_v$, have been estimated directly from the initial state of the system. The latter in particular takes a value that is in agreement with the one given by Eq. \eqref{Eff_mass}, obtained using the variational ansatz $\rho/\rho_B=x/\sqrt{2+x^2}$ \cite{stringari_book,pethick_book} for the density profile of the vortex (with $x=r/\xi$ the dimensionless coordinate), and the ratio $L/\xi\approx 32$ where $L$ is the size of the cloud. Accordingly, the value $\zeta\left(L/\xi\right)\approx 13$ has been obtained for the parameter defined in section III, which defines the effective mass of the vortex in terms of the core mass $m_{\text{core}}$. 
 
The numerical solution reproduces qualitatively the motion predicted by the variational calculation, showing trajectories for the vortex core similar to the one represented in Fig.  \ref{fig:motion}. We do not however expect a perfect match between the two approaches. The reasons for this deviation could be many. First of all, the theory developed in the previous sections refers to the ideal case of a free unbounded vortex, whose density and phase profiles are preserved in time because of the variational ansatz we used. This vortex solution is the simplest ansatz one can make, which still reproduces the dynamics approximately. In the exact numerical calculation deformations of the initially symmetric vortex core will take place. As shown in Fig.  \ref{fig:fluctuations} the current nonlinearity gives rise to an asymmetry in the effective scattering length stemming from the different direction of the current on either side of the vortex, which in turns modifies the density on the opposite sides of the vortex core, and so its dynamics.

\section{Conclusions}

In this paper we have calculated the forces acting on a vortex which is subject to a density dependent gauge potential. We identified a standard Magnus force, but also a novel force  which stems from the current nonlinearity and gives rise to a transversal force component which is of the same form as the Iordanski force. These results indicate that even if the synthetic magnetic field is zero a vortex will still experience a force due to the Galilean invariance not being fulfilled. We chose a particular laser configuration where the laser beam was incident in the 2D plane of the cloud. Other configurations are possible, in particular a symmetric situation where the synthetic gauge potential corresponds to a uniform magnetic field. Such a scenario, with a sufficiently strong synthetic magnetic field, will give rise to a vortex lattice. This lattice will be influenced by the current nonlinearity, and is likely to deviate from the standard triangular Abrikosov lattice seen in standard superfluids. It is still an open question what the resulting vortex lattice will be in the presence of current nonlinearities, and what role it plays if the quantum Hall regime is reached. 

\newpage

\section{Acknowledgements}

S.B acknowledges support from the EPSRC CM-CDT Grant No. EP/G03673X/1, M.V and P.\"O acknowledge support from EPSRC EP/J001392/1 and EP/M024636/1.

\appendix

\section{Origin of the density dependent gauge potential}

We consider a Bose-Einstein condensate of two-level atoms, where we model the collisional interactions by a zero-range pseudo potential. We assume the two internal levels are coupled by an  external laser so that, in the rotating wave approximation (RWA), the microscopic N-body Hamiltonian describing the dynamics of the system, is given by \citep{edmonds_2013a}
\begin{equation}
	H=\sum_{n=1}^{N}{\left(\frac{\mathbf{p}_n^2}{2m}+U_{af}(\mathbf{r}_n)\right)\otimes\mathbb{I}_{\mathcal{H}/n}}+\sum_{n<\ell}^N{\nu_{n,\ell}\otimes\mathbb{I}_{\mathcal{H}/\{n,\ell\}}}.
\label{Micr_Hamil}
\end{equation}
The first term in Eq. \eqref{Micr_Hamil} is the sum of the non-interacting Hamiltonians, in which the identity operators $\mathbb{I}_{\mathcal{H}/\{n,\ell,...\}}$ act on the subspace excluding the particles $n,\ell,...$. The coupling between the two internal levels $\ket{1}$ and $\ket{2}$ is given by
\begin{equation}
	U_{af}(\mathbf{r})=\frac{\hbar\Omega}{2} \begin{pmatrix}
		0 & e^{-i\phi(\mathbf{r})}\\
	 e^{i\phi(\mathbf{r})} & 0
	\end{pmatrix}
\label{Uaf^n}
\end{equation}
where $\Omega$ is the Rabi frequency characterizing the strength of the light-matter coupling, $\phi(\mathbf{r})$ is the laser phase at the atomic's position $\mathbf{r}$, and we set the laser detuning from the atomic resonance to zero for simplicity. The second term in Eq. \eqref{Micr_Hamil} represents the pairwise interaction between the particles that, in the above assumptions, has the diagonal form $\nu_{n,\ell}=\text{diag}\left[g_{11},g_{12},g_{12},g_{22}\right]\,\delta\left(\mathbf{r}_n-\mathbf{r}_\ell\right)$, with the coupling constants given by $g_{ij}={4\pi\hbar^2 a_{ij}}/{m}$ and where $a_{ij}$ are the scattering lengths relative to the three different collision channels.

We consider the limit of weakly interacting atoms, $\rho_i a_{ij}^3\ll 1$ (with $i,j=1,2$), and we make a variational ansatz by writing the many-body wavefunction $\boldsymbol{\Psi}\left(\mathbf{r}_1,\mathbf{r}_2,...\mathbf{r}_N\right)$ of the system as the symmetrized product of the single particle spinor wave function $\boldsymbol{\phi}(\mathbf{r})$, satisfying the normalization condition $\int{d^3 r\,\boldsymbol{\phi}^\dag \boldsymbol{\phi}}=1$, so that $\boldsymbol{\Psi}\left(\mathbf{r}_1,\mathbf{r}_2,...\mathbf{r}_N\right)=\prod_{i=1}^N\phi\left(\mathbf{r}_1\right)$. We introduce then the Lagrangian of the system,
\begin{equation}
	L=\prod_{i=1}^N\left(\int{d^3 r_i}\right)\left[\boldsymbol{\Psi}^\dag\left(i\hbar\partial_t-H\right)\boldsymbol{\Psi}\right].
\label{Lagrangian_many}
\end{equation}
Upon substitution of the expression given above for the many-body wave function into Eq. \eqref{Lagrangian_many}, we obtain the Lagrangian in terms of the condensate wave function $\boldsymbol{\psi}(\mathbf{r})=\sqrt{N}\boldsymbol{\phi}(\mathbf{r})$
\begin{equation}
	L_{MF}=\int{d^3 r}\left[\boldsymbol{\psi}^\dag\left(i\hbar\partial_t-H_{MF}\right)\boldsymbol{\psi}\right]
\label{Lagrangian_single}
\end{equation}
where we defined the single particle mean field Hamiltonian $H_{MF}$ as:
\begin{equation}
	H_{MF}=\frac{\mathbf{p}^2}{2m}\otimes\mathbb{I}+U_{af}(\mathbf{r})+U_{aa}+V\left(\mathbf{r}\right)
\label{MF_Hamil}
\end{equation}
in which $\mathbb{I}$ is the $2\times 2$ identity operator acting in the space of the atomic internal degrees of freedom. In Eq. \eqref{MF_Hamil} $U_{aa}$ describes the mean field collisional effects, and is given by
\begin{equation}
	U_{aa}=\frac{1}{2} \begin{pmatrix}
		\nu_1 & 0\\
	 0 & \nu_2
	\end{pmatrix}
\label{Uaa}
\end{equation}
with
\begin{align}
	\nu_1&=g_{11}\rho_1+g_{12}\rho_2 \label{nu1}\\
	\nu_2&=g_{12}\rho_1+g_{22}\rho_2 \label{nu2}
\end{align}
and where $\rho_{i}=\left|\psi_i\right|^2$ is the density of atoms in level $\ket{i}$, $i=1,2$.

Since we are working in the weakly interacting limit, the coupling energy $\hbar \Omega$ between the internal states is typically much larger than the collisional mean field shifts. This allows us to treat the meanfield interaction as a small perturbation to the atom-field coupling. To the order $\mathcal{O}(\rho_{ij} g_{ij} /\hbar\Omega)$, its eigenstates are given by
\begin{equation}
	\ket{\chi_\pm}=\ket{\chi_\pm^{(0)}}\pm\frac{\nu_1-\nu_2}{\hbar\Omega}\ket{\chi_\mp^{(0)}}
\label{IntDress}
\end{equation}
where $\ket{\chi_\pm^{(0)}}=\left(\ket{1}\pm e^{i\phi}\ket{2}\right)/\sqrt{2}$ are the so called dressed states. The interacting dressed states in Eq. \eqref{IntDress}, represent a basis for the internal Hilbert space of the atoms, so that the condensate wave function $\ket{\psi(\mathbf{r},t)}$ can be written as $\ket{\psi(\mathbf{r},t)}=\sum_{i=\{+,-\}}{}\psi_i(\mathbf{r},t)\ket{\chi_i}$.

In order to capture the dynamics of the $\pm$ component of the condensate we use the adiabatic assumption, according to which $\psi_\mp(\mathbf{r},t)\equiv 0$ (which is valid as long as the detuning induced by the collisional effect is small compared to $\hbar\Omega$), and we consider the projection of the mean field Lagrangian in Eq. \eqref{Lagrangian_single} onto the subspace spanned by the corresponding $\left(\ket{\chi_\pm}\right)$ dressed state. We obtain then the mean field Lagrangian for the relevant condensate component of the form
\begin{equation}
	L_\pm=\int{d^3 r}\left[\psi_\pm^\dag\left(i\hbar\partial_t-H_\pm\right)\psi_\pm\right]
\label{Lagrangian_pm}
\end{equation}
where
\begin{equation}
	H_\pm=\frac{\left(\mathbf{p}-\mathbf{A}_\pm\right)^2}{2m}+W\pm\frac{\hbar\Omega}{2}+\frac{g}{2}
\label{Hamil_pm}
\end{equation}
is the Hamiltonian describing the dynamics of the $\pm$ component of the condensate,
with $g=(g_{11}+g_{22}+2g_{12})/4$, while $\mathbf{A}_{\pm}=-\Melement{\chi_\pm}{\chi_\pm}{\mathbf{p}}$ and $W={\left|\Melement{\chi_+}{\chi_-}{\mathbf{p}}\right|^2}/{2m}$ are respectively the scalar and vector potential arising from the adiabatic projection of the full system onto one of the subspaces spanned by the dressed states.

Substituting Eq. \eqref{IntDress} in the expression given above for the potentials, together with $\nu_1=\rho_\pm(g_{11}+g_{12})/2$, $\nu_2=\rho_\pm(g_{22}+g_{12})/2$, obtained from Eqs.\eqref{nu1} and \eqref{nu2} in the adiabatic assumption $\left(\psi_\mp\equiv 0\right)$, the synthetic potentials are given, to the leading order, by
\begin{align}
	\mathbf{A}_\pm & =\mathbf{A}^{(0)}\pm \mathbf{a}_1 \rho_\pm(\mathbf{r})\\
	W & =\frac{\left| \mathbf{A}^{(0)} \pm\right|^2}{2m}.
\label{Potentials_def}
\end{align}
Here $\mathbf{A}^{(0)}=-\frac{\hbar}{2}\nabla\phi$ is the single particle component of the vector potential, and the vector field $\mathbf{a}_1=\nabla\phi\,(g_{11}-g_{22})/8\Omega$ controls the strength of the first order nonlinear, density dependent contribution.

By minimizing the action $S_\pm=\int{d^3\mathbf{r}\mathcal{L}_\pm}$ with respect to $\psi_\pm^*$, with the Lagrangian density defined as $\mathcal{L}_\pm\left(\psi_\pm,\psi_\pm^*\right)=\psi_\pm^*\left(i\hbar\partial_t-H_\pm\right)\psi_\pm$, we get a Gross-Pitaevskii equation for the condensate wave function, of the form
\begin{equation}
	i\hbar\frac{\partial\psi_\pm}{\partial t}=\left[\frac{\left(\mathbf{p}-\mathbf{A}_\pm\right)^2}{2m}-\mathbf{a}_1\cdot\mathbf{j}+W+g\rho_\pm\right]\psi_\pm
\label{GP}
\end{equation}
in which a current nonlinearity appears, 
\begin{equation}
	\mathbf{j}=\frac{\hbar}{2mi}\left[\psi_\pm^*\left(\nabla-\frac{i}{\hbar}\mathbf{A}_\pm\right)\psi_\pm-\psi_\pm\left(\nabla+\frac{i}{\hbar}\mathbf{A}_\pm\right)\psi_\pm^*\right].
\label{Current}
\end{equation}

\section{Transverse forces}
Here we give a more detailed description of Magnus forces in quantum fluids, outlining the basic steps needed to derive the analytic expression for the Iordanskii (or Lifshsitz and Pitaevskii) force. To this aim we consider a Bose-Einstein condensate, and we study the phonon scattering by a vortex in the hydrodynamic picture. We follow the work by Sonin \citep{Sonin_1997}.

The starting point is the Gross-Pitaevskii equation. Written in the hydrodynamic picture it reduces to the equations for the mass density $\rho=m|\psi|^2$ and the velocity field $v=\kappa/(2\pi)\nabla \phi$ where we defined the condensate wave function as $\psi=f\exp(i\phi)$:
\begin{align}
 \frac{\partial\rho}{\partial t}+\nabla\cdot\left(\rho\mathbf{v}\right)&=0\label{App_Continuity}\\
  \frac{\partial\mathbf{v}}{\partial t}+\left(\mathbf{v}\cdot\nabla\right)\mathbf{v}&=-\nabla\mu\label{App_Euler}
\end{align}
In the equations above, $\mu$ is the chemical potential, and $\kappa=h/m$ the quantum of circulation (with $m$ the mass of the atomic species). We suppose that a perturbation in the phase, of the form of a plane wave $\phi=\phi_0\exp\left(i\mathbf{k}\cdot\mathbf{r}-i\omega\,t\right)$ propagates through the condensate in the $xy$ plane, making the density and the velocity field varying in time and space. We label with $\rho_0$, $\mathbf{v}_0$ their unperturbed values, and with $\rho_1$, $\mathbf{v}_1=\kappa/(2\pi)\nabla\phi$ their periodical variations due to the sound wave, so that
\begin{align}
 \rho\left(\mathbf{r},t\right)&=\rho_0+\rho_1\left(\mathbf{r},t\right)\label{App_rho}\\
 \mathbf{v}\left(\mathbf{r},t\right)&=\mathbf{v}_0+\mathbf{v}_1 \left(\mathbf{r},t\right).\label{App_vel}
\end{align}
Furthermore, we suppose that a vortex line along the $z$ direction is present in the condensate, generating the velocity field $\mathbf{v}_0\equiv\mathbf{v}_v\left(\mathbf{r}\right)=\boldsymbol{\kappa}\times\mathbf{r}/2\pi r^2$, and moving in the $xy$ plane with the constant velocity $\mathbf{v}_L=\mathbf{v}_1(\mathbf{0},t)$, according to the Helmholtz theorem, since no external forces act on the fluid. With these assumptions, the linearised hydrodynamical equations read
\begin{align}
 &\frac{\partial\rho_1}{\partial t}+\rho_0\nabla\cdot\mathbf{v}_1=-\mathbf{v}_v\cdot\nabla\rho_1\label{App_Continuity_Lin}\\
 &\frac{\partial\mathbf{v}_1}{\partial t}+\frac{c^2}{\rho_0}\nabla\rho_1=\nabla\left[\mathbf{v}_v\cdot\mathbf{v}_1(\mathbf{r})\right]-\nabla\left[\mathbf{v}_v\cdot\mathbf{v}_1(\mathbf{0})\right]\label{App_Euler_Lin}
\end{align}
where the relation $\partial\mathbf{v}_v/\partial t=-\left(\mathbf{v}_L\cdot\nabla\right)\mathbf{v}_v$, and the vector identity $\left(\mathbf{v}\cdot\nabla\right)\mathbf{v}=\nabla\left(v^2/2\right)-\mathbf{v}\times\left(\nabla\times\mathbf{v}\right)$ have been used. From Eq. \eqref{App_Euler_Lin} we obtain the expression for the density
\begin{equation}
	\rho_1=-\frac{\rho_0}{c^2}\frac{\kappa}{2\pi}\left\{\frac{\partial\phi}{\partial t}+\mathbf{v}_v\cdot\left[\nabla\phi(\mathbf{r})-\nabla\phi(\mathbf{0})\right]\right\}
\label{App_rho1}
\end{equation}
that, substituted into Eq. \eqref{App_Continuity_Lin}, gives the equation for the phonon-induced phase
\begin{equation}
	\frac{\partial^2\phi}{\partial t^2}-c^2\nabla^2\phi=-2\mathbf{v}_v(\mathbf{r})\cdot\nabla\frac{\partial}{\partial t}\left[\phi(\mathbf{r})-\frac{1}{2}\phi(\mathbf{0})\right].
\label{App_phi}
\end{equation}
In the limit $\xi/\lambda\sim\kappa k/c \ll 1$, where $\lambda=2\pi/k$ is the wavelength of the excitation, $c$ is the sound speed, and $\xi\sim\kappa/c$ is the vortex core radius, one can treat the right-hand-side of Eq. \eqref{App_phi} as a small perturbation. Arresting the resulting Born series to the first order with respect to the perturbation parameter $\kappa k/c$, the phase is
\begin{multline}
	\phi=\phi_0\exp(-i\omega t)\left\{\exp(i\mathbf{k}\cdot\mathbf{r})+\frac{ik}{4c}\int{d\mathbf{r}'}\right.\\
	\left.\times H_0^{(1)}\left(k|\mathbf{r}-\mathbf{r}'|\right)\mathbf{k}\cdot\mathbf{v}_v(\mathbf{r}')\left[2\exp(i\mathbf{k}\cdot\mathbf{r}')-1\right]\right\}
\label{App_phi_Born}
\end{multline}
where $H_0^{(1)}$ is the zero-th order Hankel function of the first kind, and $i/4\, H_0^{(1)}\left(k|\mathbf{r}-\mathbf{r}'|\right)$ is the Green function of the two-dimensional wave equation: $\left(k^2+\nabla^2\right)\phi(\mathbf{r})=-\delta_2\left(\mathbf{r}-\mathbf{r}'\right)$. As pointed out in \cite{Sonin_1997}, the standard scattering theory fails when applied to Eq. \eqref{App_phi_Born}, since it leads to a singularity in the scattering amplitude, for small values of the scattering angle $\varphi$ between the incident wave vector $\mathbf{k}$ and the wave vector after scattering. 
This procedure would consist in looking at the scattered wave at a large distance from the scattering potential, which is assumed to be confined in a finite region, and taking advantage of the asymptotic form of the Hankel function for large values of its argument. 
This assumption is not true in our case, because of the long range character of the vortex velocity field, which slowly decays as $1/r$. An exact calculation of the integral in Eq. \eqref{App_phi_Born} is given in \cite{Sonin_1976}, and results for $\varphi\ll 1$ in an expression for the phase
\begin{equation}
	\phi=\phi_0\exp(-i\omega t)\left[1+\frac{i\kappa k}{2c}\Phi\left(\varphi\sqrt{kr/2i}\right)\right]
\label{App_phi_exact}
\end{equation}
where $\Phi(z)$ is the error function. The forces acting on the vortex can be obtained by calculating the momentum flux through a cylindrical boundary enclosing the vortex line. Since we are looking for the effect due to the phonon wave, we just consider here the relative contribution to the momentum-tensor of the fluid, which is given by
\begin{multline}
	\Pi_{ij}^{\text{ph}}=\left<P_2\right>\delta_{ij}+\left<\rho_1 v_{1i}\right>v_{0j}+\left<\rho_1 v_{1j}\right>v_{0i}\\
	+\rho_0\left<v_{1i}v_{1j}\right>.
\label{App_mom_tens}
\end{multline}
In Eq. \eqref{App_mom_tens}, $P_2$ is the second order term of the pressure with respect to the wave amplitude, where we indicated with $\left<...\right>$ average values of the fluctuating quantities. The net force is then given by the flux $\int{dS_j\Pi_j}$, with $dS_j$ the components of the outward vector normal to the circular boundary, whose magnitude is equal to the elementary area. It can be shown that only the small angle region (labelled the "interference'' region in \cite{Sonin_1997}) contributes to the momentum flow. The transverse dimension of such a region is $d\sim\sqrt{r_0/k}$, where $r_0$ is the radius of the boundary at which the momentum balance is evaluated, and corresponds to angles of the order $\varphi\sim d/r_0=1/\sqrt{k r_0}$. In this region, the component of the velocity normal to the incident wave vector $\mathbf{k}$ is equal to $v_{1\perp}=(\kappa/2\pi r)({\partial\phi}/{\partial\varphi})$ (with $\phi$ given  in Eq. \eqref{App_phi_exact}), which results in the transverse force
\begin{equation}
	\kappa j^{\text{ph}}(\mathbf{p})
\label{App_Iord}
\end{equation}
where $j^{\text{ph}}(\mathbf{p})=\left<\rho_1\mathbf{v}_1\right>=n(\mathbf{p})\mathbf{p}$ is the average mass current in the reference frame co-moving with the vortex, and $n(\mathbf{p})=\rho_0\phi_0{\kappa p}/{8\pi^2\hbar^2 c}$ is the effective number of phonons with momentum $\mathbf{p}$. In thermal equilibrium at $T>0$, the number of phonons is given by the Planck distribution $n_p(\mathbf{p})=\left[\exp\left(\epsilon(\mathbf{p})/k_B T\right)-1\right]^{-1}$, where $\epsilon(\mathbf{p})$ is the energy of the quasi-particles in the reference frame moving with their drift velocity $\mathbf{v}_n$. The total force is then given by integrating the expression given above for a single phonon wave, over all the contributions from the other modes $\int{d\mathbf{p}\,n_p(\mathbf{p})\mathbf{p}}$. This gives the expression of the Iordanskii force in terms of the thermal density $\rho_n$:
\begin{equation}
	\rho_n\left(\mathbf{v}_L-\mathbf{v}_n\right)\times\boldsymbol{\kappa}.
\label{App_Iord_tot}
\end{equation}


\end{document}